\documentclass{iau}

\usepackage{amsmath}
\usepackage{graphicx}
\usepackage{multirow}
\usepackage{natbib}
\usepackage{url}
\UseRawInputEncoding
\begin{document}

\lefttitle{Peel et al.}
\righttitle{SatHub}

\volno{385}  
\setcounter{page}{1}
\journaltitle{Astronomy and Satellite Constellations: Pathways Forward}
\editors{P. Grimley, eds.}

\doival{10.1017/xxxxx}


\title{Summary of SatHub, and the current observational status of satellite constellations}

\author{Mike~W.~Peel$^{1,2}$, Siegfried~Eggl$^{1,3}$, Meredith~Rawls$^{1,4}$, Michelle Dadighat$^{1,5}$, Piero~Benvenuti$^{1,6}$, Federico~di~Vruno$^{1,7}$, Connie~Walker$^{1,5}$}
\affiliation{1. IAU Centre for the Protection of the Dark and Quiet Sky from Satellite Constellation Interference, France. \email{sathub@cps.iau.org}}
\affiliation{2. Imperial College London, Blackett Lab, Prince Consort Road, London SW7 2AZ, UK.}
\affiliation{3. Department of Aerospace Engineering/Department of Astronomy, University of Illinois at Urbana-Champaign, Illinois, USA.}
\affiliation{4. Department of Astronomy/DiRAC/Vera C. Rubin Observatory, University of Washington, Seattle, Washington, USA.}
\affiliation{5. NSFs NOIRLab, Tucson, Arizona, USA.}
\affiliation{6. Emeritus, Department of Astronomy, University of Padova, Italy.}
\affiliation{7. SKA Observatory, Jodrell Bank, Macclesfield SK11 9FT, United Kingdom.}

\begin{abstract}
SatHub is one of the four hubs of the IAU Centre for the Protection of the Dark and Quiet Sky from Satellite Constellation Interference (CPS). It focuses on observations, data analysis, software, and training materials to improve our understanding of the impact of satellite constellations on astronomy and observers worldwide. As a preface to more in-depth IAUS385 sessions, we gave a summary of some recent work by SatHub members and the current status of satellite constellations, including optical and radio observations. We shared how the audience can join or get more involved, e.g., via the CPS Slack for asynchronous collaboration. We also touched on what a future with hundreds of thousands of constellation satellites might look like.
\end{abstract}


\maketitle

\section{Introduction}
\begin{figure}[h]
  \center
  \includegraphics[width=0.9\textwidth]{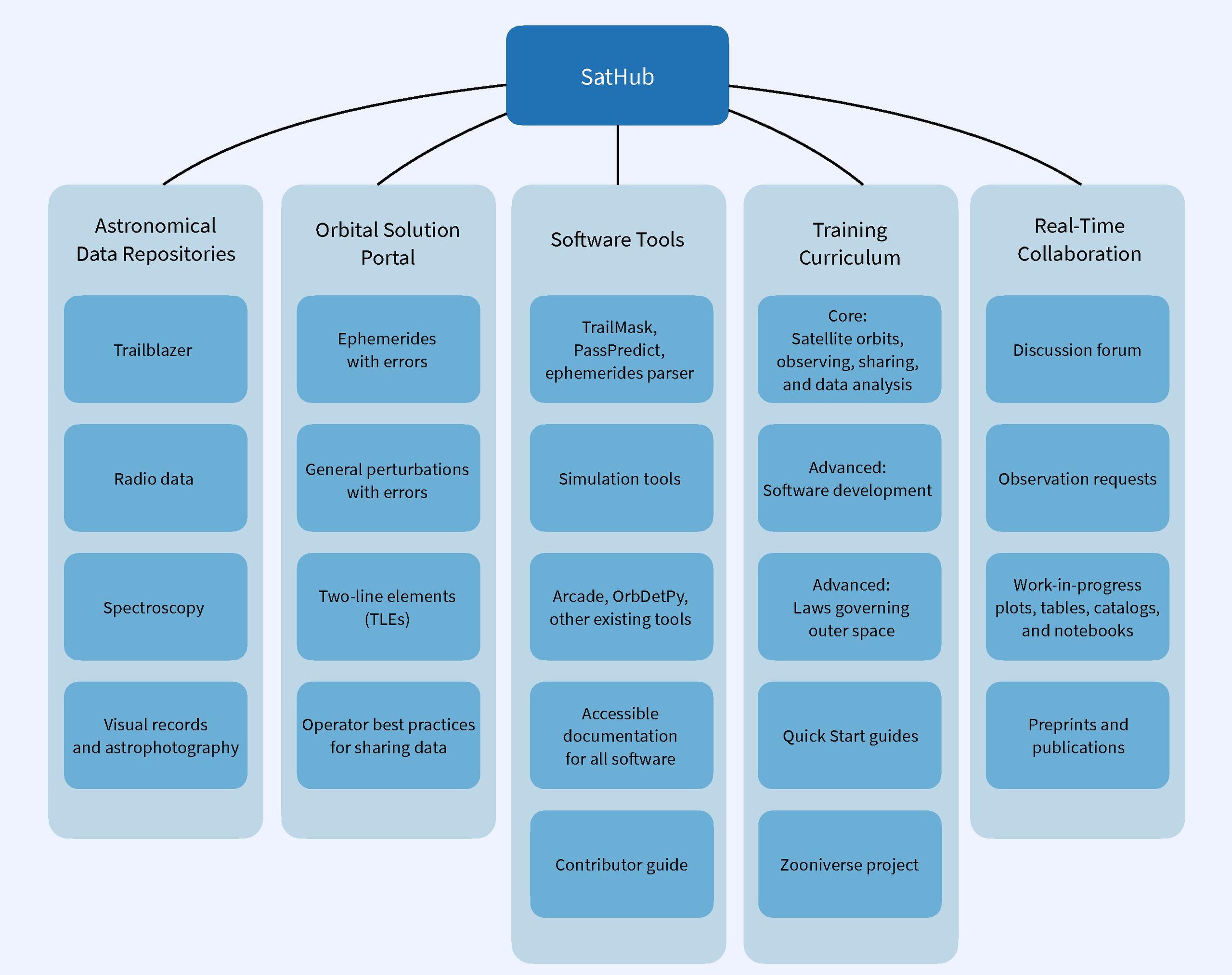}
  \caption{SatHub outline from the SATCON2 Observations Working Group \citep{rawls_satcon2_2021}.}
  \label{fig:satcon2}
\end{figure}

SatHub is part of the International Astronomical Union (IAU) Centre for the Protection of the Dark and Quiet Sky from Satellite Constellation Interference (CPS). It is a public, community driven, coordinated hub focused on the collection and analysis of artificial satellite observations, as well as carrying out training and outreach related to satellite constellations. It was formed following the recommendations from SATCON2 \citep{rawls_satcon2_2021}, and adopted the outline structure established in that report, see Fig.~\ref{fig:satcon2}. SatHub is still under construction, and has built-in flexibility to change and evolve based on community feedback and needs.

SatHub is co-led by three people: Meredith Rawls (University of Washington, WA, USA; remote attendee of IAUS385); Mike Peel (Imperial College London, UK; in-person attendee); and Siegfried Eggl (University of Illinois at Urbana-Champaign, USA; in-person attendee). In addition, Michelle Dadighat (NOIRLab, in-person attendee) is a software engineer working on SatHub software development.

\section{Software development}
A central part of the work of SatHub involves software development to make it easier to identify, avoid, and remove satellites from astronomical observations across the electromagnetic spectrum. Software development is actively underway, and we have marked in the section below where things are already working well / up-and-running as ``done''; in development as ``developing''; and still in need of doing as ``to be done''.

\subsection{Data repository and exchange}
One of the primary purposes of SatHub is to serve as a Hub for Data Exchange from brightness measurements. This will be a central place where data can be deposited and exchanged with others. Such a system needs to be publicly available, easily accessible, user-friendly, and well documented.

The repository will enable the submission and retrieval of satellite brightness measurements (made visually by eye, observed with optical telescopes, etc.; developing); collections of optical/near-IR images with satellite streaks (developing); spectra contaminated with reflected solar spectrum (to be done); and radio data affected by satellite interference (to be done).

We already welcome contributions from everyone who is willing to share observing campaign plans and data. We aim to have easy-to-use interfaces for both professional and amateur astronomers to upload and download satellite-affected images (developing). Software developers/contributors are welcome to contact us to help with this work!

\subsection{Orbital solutions portal}
A critical need for astronomy is high accuracy predictions of where satellites will be seen from any observatory around the world, which we aim to enable through the Orbital Solution Portal. This will provide public, standardised access to orbital solutions of artificial satellites. NOIRlab has recently hired a software developer (Michelle Dadighat) to work with SatHub on this and other tools.

The portal will convert general perturbation-style two-line element (``TLE'') solutions to on-sky positions, with automatic synchronisation with complementary services and databases (done). It is updated every 8 hours for Starlink satellites, with others updated every 24 hours (due to the availability of published TLEs). It will include error bars with all orbital solutions (developing), and have built-in robustness by using multiple backend systems to provide TLEs and position predictions (CelesTrak\footnote{\url{https://celestrak.org/}}, Privateer\footnote{\url{https://www.privateer.com/}}, etc.; developing). Ultimately it will include the full history of all TLEs, so positions can be predicted for any satellite at any time, which is critical when analysing archival and long-duration survey data (to be done; funding applied for).

At present we have a prototype SatChecker that does some of this\footnote{\url{https://satchecker.readthedocs.io}}. We welcome contributions to this work. In particular, we need industry to cooperate and share the highest accuracy orbital information about their satellites in a timely, regular manner. We also need astronomers to check that orbits are accurate by observing and measuring satellites, and reporting inaccuracies back to us and to industry.

\subsection{Other software tools}

In addition to the two key tools above, we are also working on other tools, as described in the SATCON2 report \citep{rawls_satcon2_2021}. We aim to make sure that all software tools have user-friendly documentation, support, and are actively maintained (developing). In addition, we plan to have a standard test suite supporting a wide range of instrument and satellite signature properties to support software development (to be done).

We encourage contributions to open-source software development, particularly through the tools developed within CPS. We also want to test the software with data from many different telescopes and cameras. We are looking for funding to work on software development.

In addition to software developed within CPS, we are also seeking to understand existing tools and their availability, so that we can reduce duplication and avoid recreating what already exists. SatHub is running a software survey\footnote{\url{https://forms.gle/FzkT1FekoVb2Fxqi6}}, with one submission per tool requested. Tools inputted into this survey before and during IAUS385 were then discussed and worked on at the SatHub hackathon on the last day of the symposium.

\section{Real-time collaboration}
SatHub is working on establishing communication and collaboration across the astronomy community about satellite constellations in an open and collaborative way. We mostly do this through the IAU CPS Slack, which all members of CPS can access, with the main {\tt \#sathub} channel and a host of sub-channels on specific topics (for example, {\tt \#sathub\_optical\_observing}).

We also have a SatHub GitHub organisation\footnote{\url{https://github.com/iausathub}}, which anyone can use to share open-source code, and we encourage documentation that follows ``Write the Docs''\footnote{\url{https://www.writethedocs.org/}} best practices (developing). We plan to have a website that will direct to new and existing resources (to be done).

We offer a paper review process, which provides free peer reviews within two weeks for satellite-related papers. The aim is to provide feedback from peers who are also working on satellites within CPS, to ensure accuracy and connections between publications produced by different groups. The reviews are carried out by CPS members, for CPS members, and while being a member is not strictly required to use this service, we encourage everyone interested in the topic to join CPS. Papers that have gone through this peer review process can include CPS as an affiliation and acknowledgement. A recent example of a paper that has gone through the process is \citet{2023arXiv230914152M}. We are currently working on documenting the process, to be published soon on the CPS website.

\section{Potential impact at radio frequencies}
As well as optical reflections, satellites actively transmit at radio frequencies. We don't yet have a complete picture of the impact that radio transmissions from satellite constellations will have across the full frequency range used by radio astronomy, and we need observations to assess actual impact. Various talks at IAUS385 covered this topic, and are described later in this Proceeding.

\begin{figure}[h]
  \center
  \includegraphics[width=0.4\textwidth,trim={0cm 0cm 0.5cm 0cm},clip]{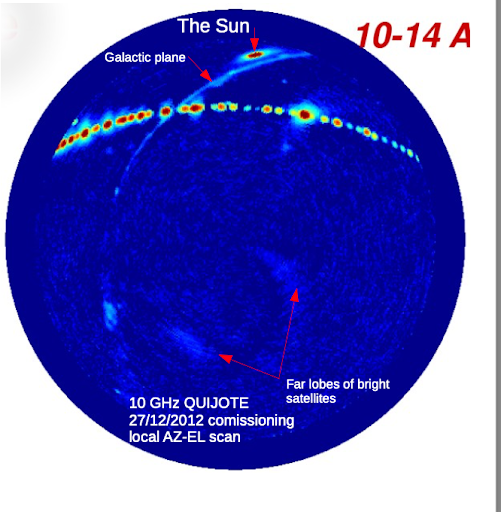}
  \caption{QUIJOTE local map of the radio sky above Tenerife, with relevant features highlighted. Geostationary satellites appear as a band across the sky, similar in brightness to the Sun (top of the image), and also appearing in telescope sidelobes in the bottom half of the image.}
  \label{fig:quijote}
\end{figure}

Satellite constellations actively transmit at radio frequencies to provide internet connectivity services. This is currently at $\sim$10--20\,GHz (12.7--14.7\,GHz for Starlink, others use different frequencies), various satellite constellations are looking at using higher frequencies for higher bandwidth in the near future ($\sim$40--200\,GHz). Typically there are also octaves of radio emission at, e.g., 2$\times$ the base frequency (or even 3, 5, 10$\times$ etc.), which can also impact radio astronomy since the main signals are so bright. In addition, telescopes are not only sensitive to the direction they are observing, but also to other directions through sidelobes (reflections from different parts of the telescope other than the main mirror). If these couple to satellite transmissions, this can be a concern, particularly for CMB experiments who are mapping large angular scale emission and could see interference over large parts of the sky away from the satellite positions. This can be seen in data from the QUIJOTE MFI instrument \citep[Génova-Santos in prep.]{2010ASSP...14..127R,2023MNRAS.519.3383R} in 2012 observing at 10--14\,GHz (Fig.~\ref{fig:quijote}), where geostationary satellites appear as bright as the Sun, and sidelobes were also initially visible---these were later avoided by adding additional baffles around the telescope aperture.

At lower frequencies ($\sim$100-200\,MHz), unintended emission has been observed by LOFAR \citep{2023A&A...676A..75D} and SKA low \citep{2023A&A...678L...6G}. This emission comes from the electronics on board the satellites, and was not foreseen by the satellite operators before being detected by radio telescopes.

The types of detectors used in radio telescopes, particularly at high frequencies, can make it difficult to filter out satellite emission. Radiometers require FPGAs rather than broad-band detectors to be able to excise frequencies affected by interference - and these are much more expensive than simple detectors. At high frequencies ($>$30\,GHz), superconducting detectors like Transition Edge Sensors (TESs) and Kinetic Inductance Devices (KIDs) are inherently broad-band and can only have optical filters to remove unwanted emissions - mainly high and low-pass filters that define the overall bandwidth. In such systems, interference can only be removed by flagging the time-ordered data.

\begin{figure}[h]
  \center
  \includegraphics[width=0.9\textwidth]{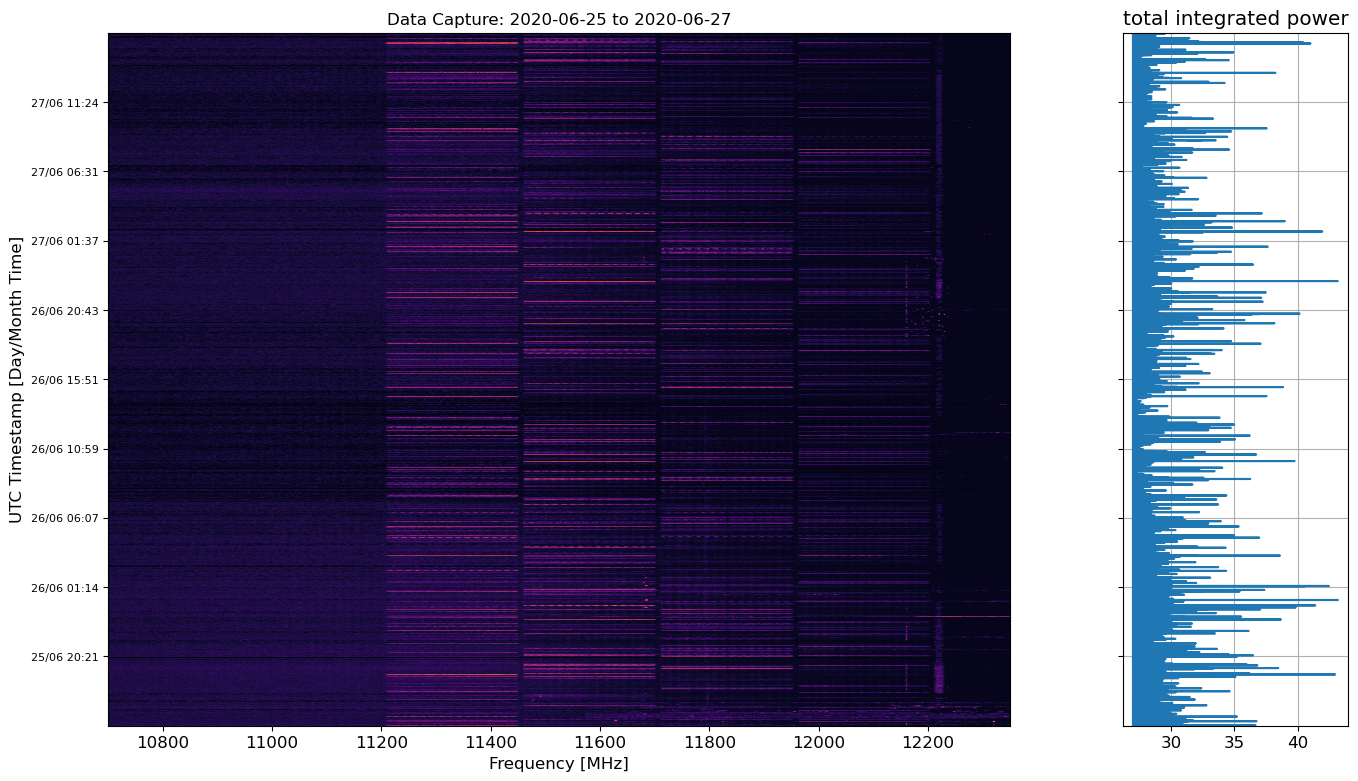}
  \caption{Observations of Starlink with a satellite dish in 2022, demonstrating broad bandwidth and high variability, courtesy of F. Di Vruno.}
  \label{fig:variable}
\end{figure}

In addition, satellite emission can be highly variable (see Fig.~\ref{fig:variable}), both because of their high angular velocities, and because they can be changing transmission direction, e.g., using phased arrays. It is important to accurately know satellite positions, otherwise such emission could be mis-identified as transient astronomical signals.

\section{Observing campaigns}
SatHub runs observing campaigns for specific high-importance satellite constellation launches, such as a the launch of a new type or generation of satellite. A strength of SatHub is our ability to recruit observers from across the globe with a wide variety of expertise to assemble a more complete picture of how a satellite or constellation of satellites is affecting the sky.

\subsection{BlueWalker3}

\begin{figure}[h]
  \center
  \includegraphics[width=0.6\textwidth]{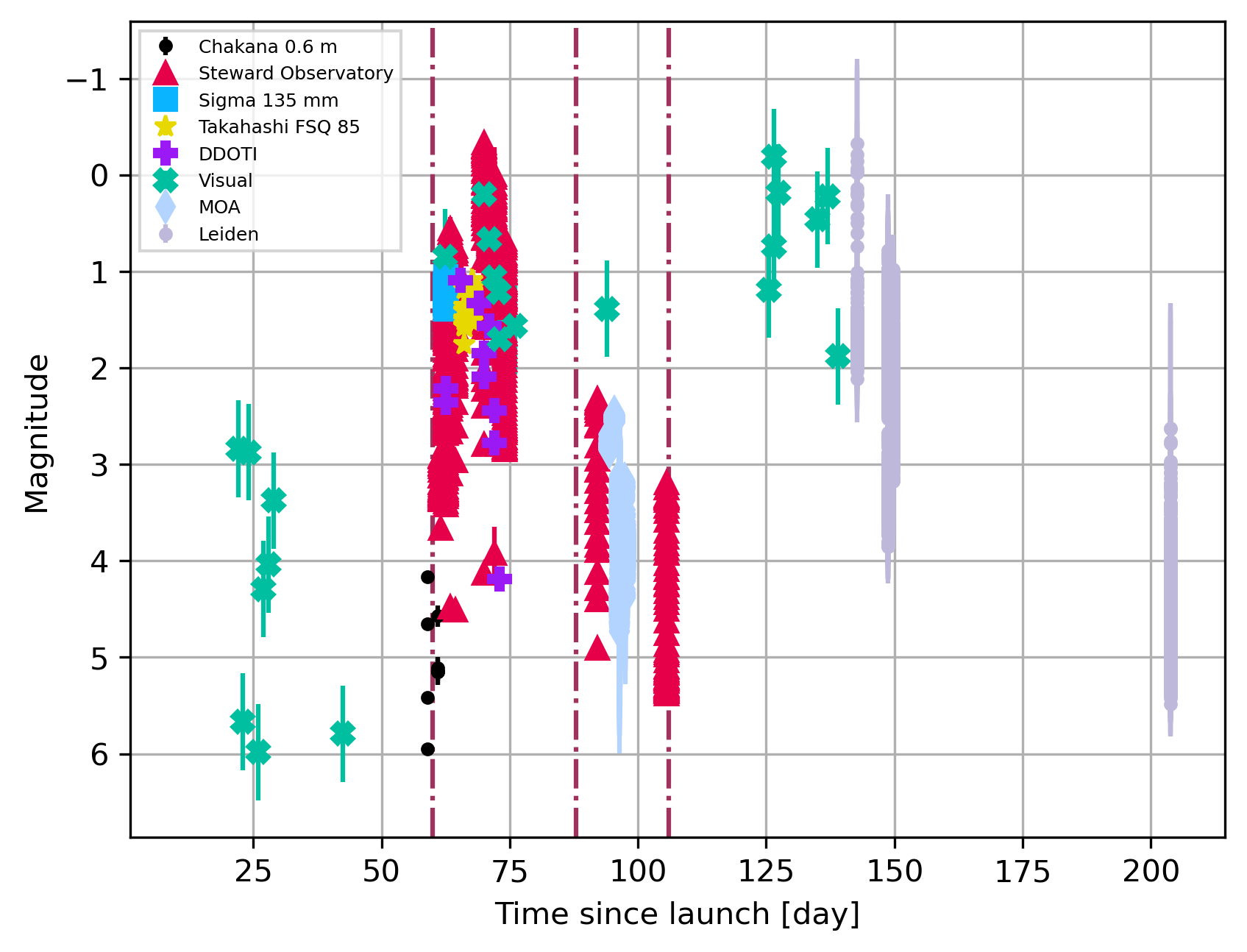}
  \caption{Brightness of BW3 over time, from \citet{nandakumar_high_2023}. The first dashed vertical line shows the point where BW3's phased array was deployed and the satellite became exceptionally bright.}
  \label{fig:bw3_brightness}
\end{figure}

\begin{figure}[h]
  \center
  \includegraphics[width=0.5\textwidth,trim={1cm 4cm 1cm 4cm},clip]{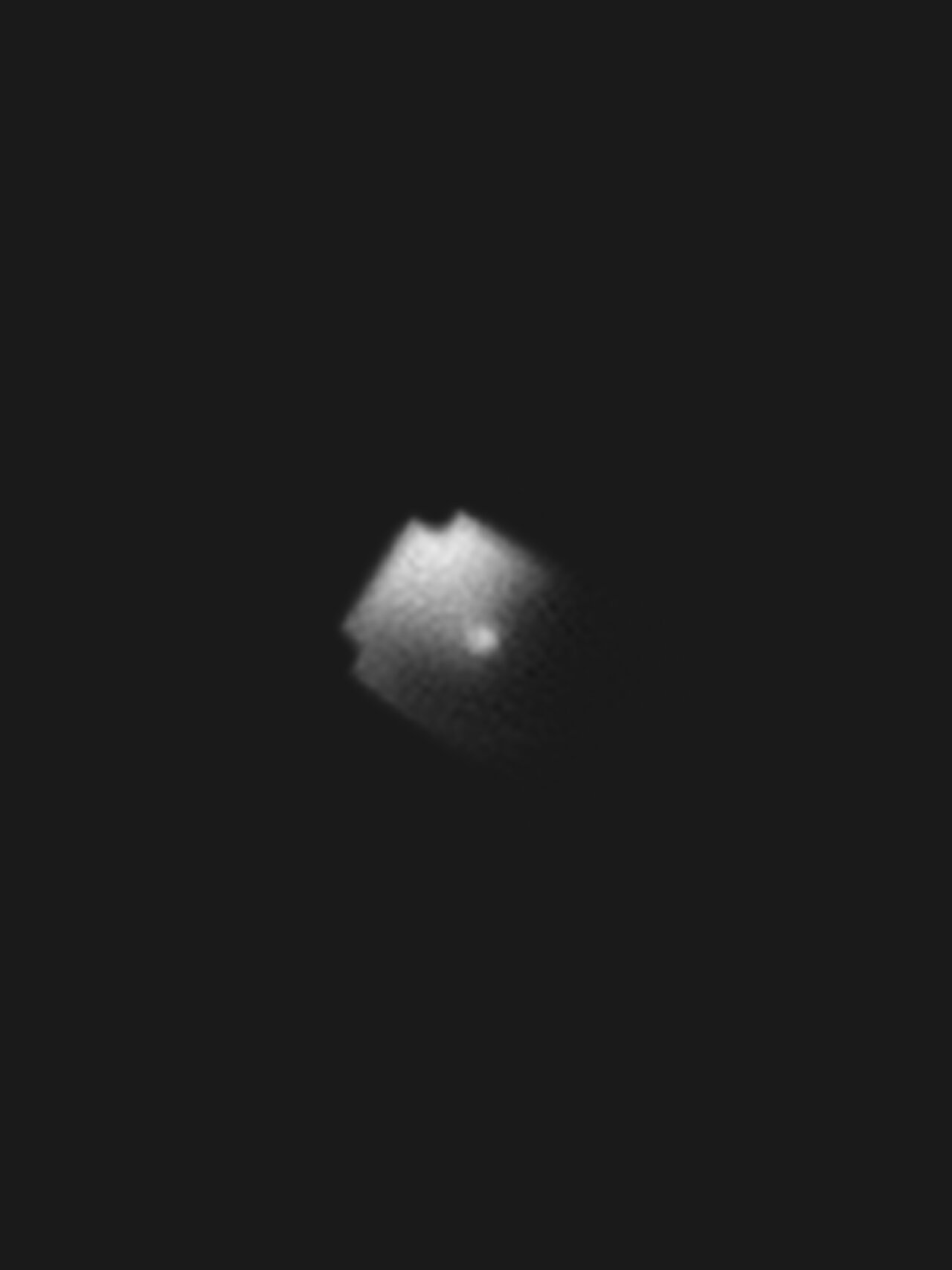}
  \caption{Image of BlueWalker 3 in orbit by M. Tzukran (CC-BY-4.0)}
  \label{fig:bw3_orbit}
\end{figure}

AST BlueWalker 3 is a prototype satellite that was launched in 2022 to test direct connectivity between satellites and unmodified mobile phones. To do this, it uses a 64\,m$^2$ phased array that was folded on launch and deployed in orbit. Already bright after launch, the satellite became one of the brightest objects in the sky once unfolded (Fig.~\ref{fig:bw3_brightness}), and is also large enough to be imaged from the ground (Fig.~\ref{fig:bw3_orbit} \footnote{\url{https://www.astrobin.com/tffts1/}}
  \footnote{\url{https://noirlab.edu/public/images/ann23027b/}}
  ). The launch vehicle adapter was also bright, and initially untracked with public TLEs. In addition, we demonstrated that the position accuracy decays between releases of TLEs, highlighting the importance of regularly releasing updated TLEs.

The observing campaign was published in Nature \citep{nandakumar_high_2023}, with an online announcement\footnote{\url{https://noirlab.edu/public/announcements/ann23027/}}. The announcement included a video of BW3, as well as Starlink satellites, as they transited the sky---showing how bright BW3 was compared with Starlink satellites and nearby stars.\footnote{\url{https://noirlab.edu/public/videos/ann23027a/}}

\subsection{Ongoing}

Starlink started launching its Generation 2 mini satellites in 2023. These have optical mitigations using Bragg mirrors\footnote{\url{https://api.starlink.com/public-files/BrightnessMitigationBestPracticesSatelliteOperators.pdf}}, and CPS seeks to quantify how well these mitigations have worked. Early observations indicate mixed results so far, with some satellites being faint, but some remaining bright. This is concerning as the full Generation 2 satellites will be bigger than the minis (which were downsized to fit into the Falcon 9 launcher), and potentially will be correspondingly brighter. The results of the observing campaign are planned to be published in early 2024.

Observations are also continuing of BlueWalker 3, using SCUBA2/JCMT to observe the satellite (and also the International Space Station) at submm wavelengths, to see how thermally bright the satellites are. The observations were recently approved, and will start shortly; if successful these will lead to future observations of smaller satellites such as Starlink and OneWeb.

We are also preparing for other observing campaigns in the future, for example, of the Amazon Kuiper satellites. Participation in, and ideas for, observing campaigns are very welcome.

\section{Get involved}
SatHub needs the help of everyone interested in the impact of satellite constellations on astronomy. We are seeking sky observers, data analysts, software developers, industry experts, students, and others to help with this work. Additionally, as the satellite population changes, the evolving impacts that this will have on astronomy require ongoing observer-operator dialogs. Information in SatHub will be public, open, and accessible to support real-time collaboration. We aim to join innovation with existing solutions, prioritize ease of use, and enable coordination among multiple stakeholders.

To contribute to SatHub, first of all, join CPS! After that, join the SatHub channel at the CPS Slack, which is where most of the discussions take place. Please also share your observations, experience, relevant code and work plans with us, and contribute to open-source software and documentation development! There are no requirements, prerequisites, or fees for membership---you just have to follow our collaboration agreement\footnote{\url{https://docs.google.com/forms/d/e/1FAIpQLSe6w5rZhxvF0ozGe1dMssLTolua1Apq1_QjJ8aLWprGUGQGog/viewform}} and our code of conduct\footnote{\url{https://tinyurl.com/IAUCPSCoC}}. You can apply for membership at \url{https://cps.iau.org/}.

\section*{Acknowledgements}
MP, SE, MD acknowledge travel funding from IAU CPS to attend IAUS385.

\bibliographystyle{iaulike}
\bibliography{refs}

\end{document}